\documentclass[aps,prb,reprint,superscriptaddress,nofootinbib,longbibliography,floatfix]{revtex4-2}
\usepackage{amsmath,amssymb,bm}
\usepackage{graphicx}
\usepackage{colordvi}
\usepackage{mathrsfs}
\usepackage{verbatim}
\usepackage{dcolumn}
\usepackage{epsfig}
\usepackage{subfigure}
\usepackage{makecell}
\usepackage{float}
\usepackage{esint}
\usepackage{bbding}
\usepackage{siunitx}
\usepackage{pifont}
\usepackage[colorlinks=true,linkcolor=blue,citecolor=blue,urlcolor=blue]{hyperref}

\usepackage{ulem}

\begin{document}

\title{Quasi-two-dimensional Majorana zero modes from finite-size-coupled chiral hinge states}

\author{Wenhao Liang}
\affiliation{Department of Physics, The Hong Kong University of Science and Technology, Clear Water Bay, Hong Kong, China}

\author{Xun-Jiang Luo}
\email{xjluo@hmfl.ac.cn}
\affiliation{Anhui Province Key Laboratory of Low-Energy Quantum Materials and Devices, 
High Magnetic Field Laboratory, HFIPS, Chinese Academy of Sciences, Hefei, Anhui 
230031, China }
\date{\today}

\begin{abstract}
Majorana zero modes (MZMs) in topological superconductors have attracted broad research interest for their potential applications in topological quantum computation. In this work, we propose a quasi-two-dimensional route to realize spatially separated MZMs in a chiral higher-order topological insulator (HOTI) proximitized by a conventional $s$-wave superconductor through a theoretical model study. In three dimensions, the chiral HOTI hosts gapless hinge states along the $z$ direction, arising from a mass term that anisotropically gaps the surface Dirac cones of a topological insulator. By confining the sample along the $x$ direction while keeping it extended along $y$ and finite along $z$, opposite $z$-directed chiral hinge states hybridize and effectively form one-dimensional helical channels. Incorporating the superconducting proximity effect into this quasi-two-dimensional system induces effective $p$-wave pairing in these helical channels, thereby opening a topological gap. A fully open-boundary sample then hosts four localized MZMs, one at each endpoint of the helical channels, realizing a second-order topological superconductor characterized by Majorana corner modes. In addition to MZMs, we also find that superconducting pairing in this model produces extended Majorana hinge modes in three dimensions. Furthermore, representative disorder calculations indicate that these Majorana corner modes are robust against weak-to-moderate disorder, provided the excitation gap remains open. These results establish finite-size-coupled chiral hinge states as a promising platform for engineering multiple MZMs via conventional superconducting proximity effect.
	\end{abstract}

\maketitle

\section{Introduction}
\label{sec:introduction}

Majorana zero modes (MZMs) are self-conjugate excitations of topological superconductors (TSCs). Their nonlocal fermion parity and non-Abelian exchange statistics make them central objects in the realization of fault-tolerant topological quantum computation~\cite{Kitaev2001,Nayak2008RMP,Alicea2012RPP,Beenakker2013ARCMP,DasSarma2015NPJQI,SatoAndo2017RPP,Flensberg2021NatRevMater}. The minimal theoretical setting for TSCs is a spinless $p$-wave superconductor, represented by one-dimensional Kitaev chain ~\cite{Kitaev2001}. In real materials, however, intrinsic $p$-wave superconductivity remains elusive and is notoriously difficult to identify and control~\cite{MackenzieMaeno2003,SatoAndo2017RPP,Flensberg2021NatRevMater}. This has motivated intense efforts to engineer TSCs by proximitizing conventional $s$-wave superconductors to spin-orbital-coupled systems, including topological-insulator, semiconductor nanowires, and quantum-anomalous-Hall (QAH) insulators~\cite{FuKane2008PRL,Sau2010PRL,Lutchyn2010PRL,Oreg2010PRL,Mourik2012Science,NadjPerge2014Science,Lutchyn2018NatRevMater,Deng2018Nonlocality,Wang2018FeTeSe,Machida2019NatMater,Wang2015QAHSC,Chen2018QAHMajorana,Zeng2018}. In proximitized QAH systems, the interface typically hosts chiral Majorana edge modes; however, these one-dimensional propagating modes are inherently delocalized and thus unsuitable for topological quantum computation. To overcome this limitation, a promising strategy employs narrow QAH strips: finite-size coupling between opposite chiral edges reconstructs the boundary spectrum into a quasi-one-dimensional helical channel, which can be gapped by proximity-induced pairing to yield localized MZMs~\cite{Chen2018QAHMajorana,Zeng2018}. This approach is experimentally feasible and provides a practical route to engineering localized MZMs.

In recent years, higher-order topological phases have emerged as a fertile ground for realizing novel boundary states \cite{ShiozakiSato2014PRB,Benalcazar2017Science,Benalcazar2017PRB,Langbehn2017PRL,Song2017PRL,Schindler2018SciAdv,Khalaf2018PRB,Geier2018PRB,PhysRevB.107.045118,PhysRevB.108.075143}. In a $d$-dimensional $n$th-order topological phase, gapless boundary modes are confined to $(d-n)$-dimensional boundaries. For instance, a three-dimensional second-order topological insulator is characterized by fully gapped bulk and surface spectra but hosts gapless one-dimensional hinge states. When time-reversal symmetry is broken, the system becomes a chiral higher-order topological insulator (HOTI) with chiral hinge states reminiscent of the chiral edge states in QAH  insulators~\cite{Schindler2018SciAdv,Fu2021PRResearch}. Such chiral hinge states have been predicted in magnetic topological insulators and axion insulators, including MnBi$_{2n}$Te$_{3n+1}$~\cite{PhysRevLett.124.136407}, antiferromagnetic axion insulators~\cite{PhysRevLett.122.256402}, and Sm-doped Bi$_2$Se$_3$~\cite{Yue2019}.
Higher-order topological superconductors generalize this boundary hierarchy to superconducting systems, where Majorana corner or hinge modes have been proposed in various platforms~\cite{WangLinHughes2018WeakPairing,Hsu2018HOTIMajorana,Yan2018HighTPlatform,Wang2018HighTempCorner,Yan2019SOTISCHeterostructure,Ghorashi2019SecondOrderDirac,Ghorashi2020VortexSurfaceHOTI,Pan2019LatticeSOTSC,Zhu2019SecondOrderTSC,Wu2020ZeemanSOTSC,Kheirkhah2020CornerFlatBands,ZhangColeDasSarma2019PRL,Ghosh2021HierarchyHOTSC,Chew2023TBGHOTSC,Nakai2023HOTSCMultilayer,Luo2021WeakTIHOTSC,Pan2022BraidingMCM,Luo2024SpinBott}. This raises a natural question: can the chiral hinge states already present in a HOTI be leveraged to realize MZMs through a superconducting proximity structure combined with finite-size coupling?

\begin{figure}
	\includegraphics[width=8.5cm]{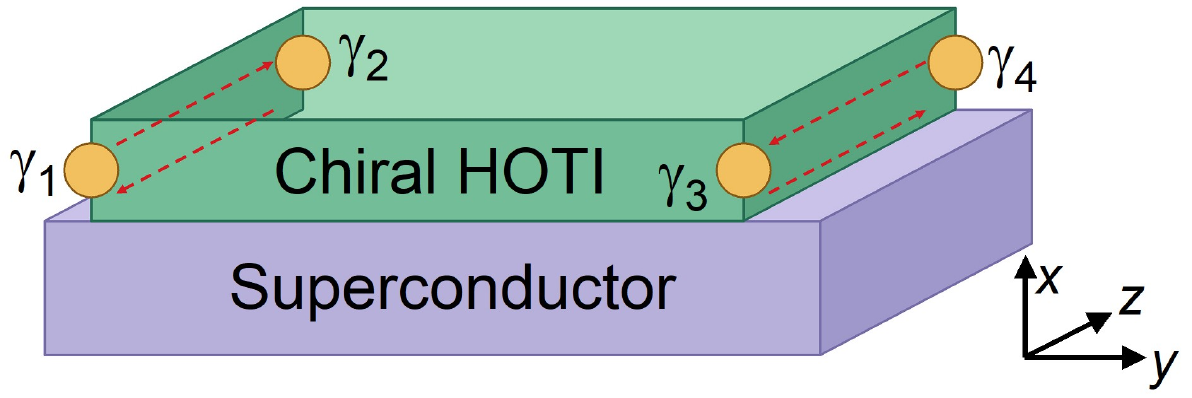}
	\caption{Schematic of the quasi-two-dimensional HOTI-superconductor geometry. A chiral HOTI slab is proximitized by a conventional $s$-wave superconductor. The sample is thin along $x$, extended along $y$, and finite along $z$. Opposite $z$-directed hinge states hybridize across the thin direction. The proximity-induced pairing gaps the two separated hinge-derived channels indicated by dashed red arrows. The four MZMs $\gamma_{1,2,3,4}$ are localized at the endpoints of these channels in the $y$-$z$ plane.}
	\label{schematic}
\end{figure}

In this work, we provide an affirmative answer by
studying a theoretical model. We introduce a chiral HOTI realized by gapping the surface Dirac cones of a topological insulator via a time-reversal-symmetry-breaking mass term, yielding chiral hinge states along the $z$ direction. When the sample is made thin along $x$ while remaining wide along $y$ and finite along $z$, opposite $z$-directed hinge states hybridize across the thin direction and effectively form two spatially separated helical channels, as sketched in Fig.~\ref{schematic}. Upon proximitizing this quasi-two-dimensional geometry with a conventional $s$-wave superconductor, each channel behaves as an effective one-dimensional $p$-wave superconducting wire. Opening the boundaries along $z$ then produces four localized MZMs, one at each endpoint of the two channels, thereby realizing Majorana corner states. We demonstrate this physical picture through numerical calculations of the spectra, real-space state densities, and the phase diagram in the model parameter space. 
Besides the localized MZMs, the considered system also exhibits
orientation-dependent chiral hinge Majorana modes in three dimensions, which serve as a
complementary signature of the superconducting HOTI. 
Furthermore, representative disorder calculations show that the Majorana corner modes are robust against weak-to-moderate disorder as long as the excitation gap remains open. Our proposal is conceptually analogous to the finite-width QAH mechanism \cite{Chen2018QAHMajorana,Zeng2018}, yet it leverages the higher-order boundary structure of a three-dimensional HOTI and is experimentally feasible.

The remainder of this paper is organized as follows. Section~\ref{sec:model} introduces the model Hamiltonian of the chiral HOTI and the surface dowain wall mechanism responsible for the chiral hinge states. Section~\ref{sec:hinge-majorana} incorporates $s$-wave pairing and investigates the orientation-dependent Majorana hinge modes in the considered superconducting system. Section~\ref{sec:q2d-mzm} presents the finite-size route to realizing four localized MZMs and the phase diagram in the model-parameter space. Section~\ref{sec:disorder} examines the robustness of the MZMs against disorders. Section~\ref{sec:conclusion} presents a brief summary and discussion. Appendices~\ref{appendixa} and \ref{app:superconducting-spectra} provide additional details complementing the main text.

\section{Chiral HOTI model}
\label{sec:model}

\begin{figure}
	\includegraphics[width=8.5cm]{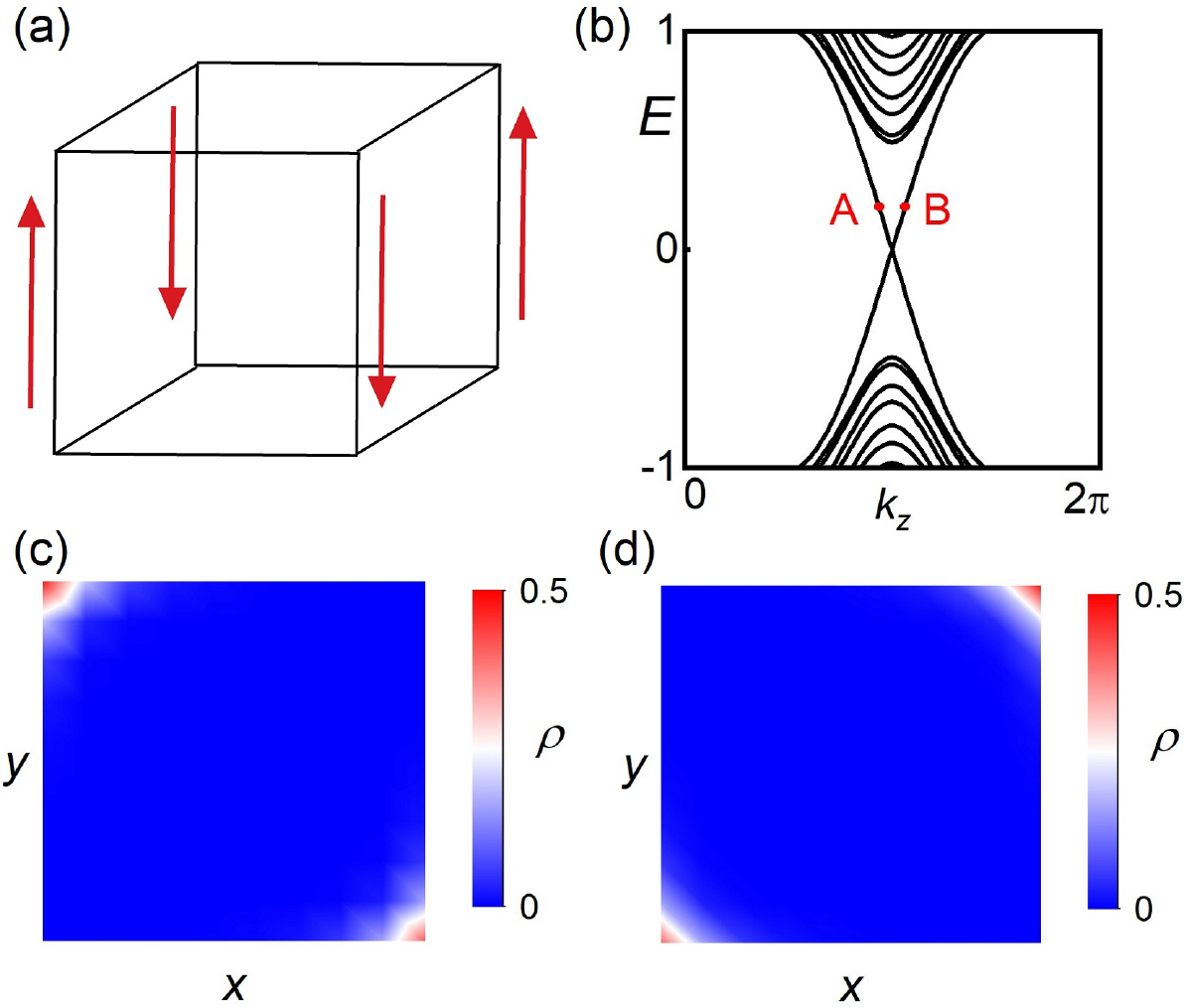}
	\caption{Energy spectrum and real space distribution of chiral hinge states. (a) Propagation directions of the $z$-directed chiral hinge states. (b) Spectrum of a $z$-periodic wire with a $10\times10$ cross section and open boundary conditions in the $x$-$y$ plane.  (c),(d) Probability densities of the states marked \textit{A} and \textit{B} in (b), showing localization on complementary diagonal hinge pairs. The model parameters are $t=1$, $M=2$, $\lambda=1$, and $m=0.5$}
	\label{HOTI_kz}
\end{figure}

\begin{figure*}
	\includegraphics[width=16cm]{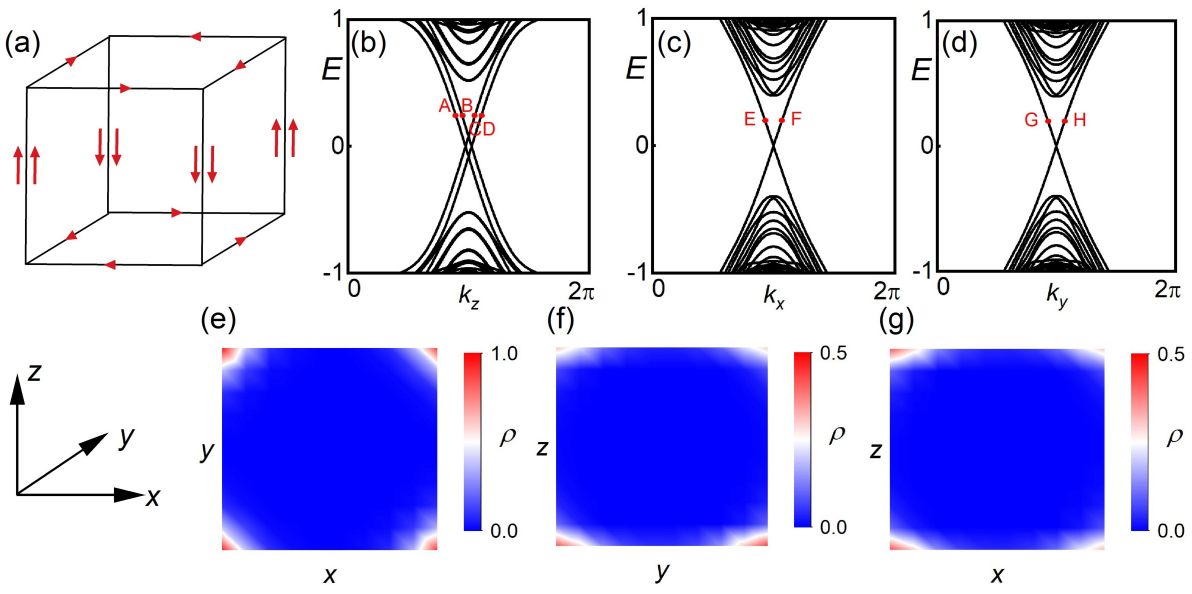}
	\caption{The energy spectrum of $H_{\text{BdG}}$ with a wire geometry and distribution of chiral Majorana hinge modes.
     (a) Schematic illustration of the flow pattern of chiral Majorana hinge modes. (b)--(d) BdG spectra for wire geometries periodic along $z$, $x$, and $y$, respectively, with a $10\times10$ cross section and open boundary conditions in the two transverse directions. Red points \textit{A}--\textit{H} mark representative in-gap hinge states. (e)--(g) Summed probability densities of the marked states in the corresponding cross sections: (e) states \textit{A}--\textit{D} in the $x$-$y$ plane, (f) states \textit{E}--\textit{F} in the $y$-$z$ plane, and (g) states \textit{G}--\textit{H} in the $x$-$z$ plane. The model parameters are $t=1$, $M=2$, $\lambda=1$, $m=1$, $\mu=0.1$, and $\Delta=0.3$.}
	\label{hinge_HOTSC}
\end{figure*}

We start from a four-band tight-binding model introduced to realize a chiral HOTI with gapless hinge states~\cite{Schindler2018SciAdv,Benalcazar2017PRB,Langbehn2017PRL,Song2017PRL,Khalaf2018PRB,Geier2018PRB,Fu2021PRResearch}. The Bloch Hamiltonian in momentum space takes the form
\begin{align}\label{eq_HOTI}
{H}(\mathbf{k})&=\left(M+t\sum_{i}\cos k_i\right)\tau_z\sigma_0+\lambda\sum_{i}\sin k_i\,\tau_x\sigma_i\nonumber\\
&+m(\cos k_x-\cos k_y)\tau_y\sigma_0,
\end{align}
where $i=x,y,z$, and Pauli matrices $\tau_i$ and $\sigma_i$ act in orbital and spin spaces, respectively. The first three terms constitute a standard Dirac model for a three-dimensional topological insulator, hosting a gapless Dirac cone on each surface. The last term denotes an anisotropic mass term that gaps the surface states and realizes higher-order topology.

We note that the model Hamiltonian preserves the combined four-fold rotation around the $z$ axis and time-reversal symmetry $\mathcal{C}_{4z}\mathcal{T}$,
\begin{align}
 (\mathcal{C}_{4z}\mathcal{T})H(k_x,k_y,k_z)(\mathcal{C}_{4z}\mathcal{T})^{-1}=H(k_y,-k_x,-k_z),
 \label{eq:C4T-real-space}
\end{align}
where time reversal is represented by $\mathcal{T}=-\tau_0\mathrm{i}\sigma_y\mathcal{K}$ and fourfold rotation by $\mathcal{C}_{4z}=\tau_0\exp(-\mathrm{i}\pi\sigma_z/4)$. This symmetry enforces an alternating sign of the mass on the side facets: the $x$- and $y$-normal surfaces acquire opposite Dirac masses for the surface Dirac cones of the topological insulator. Consequently, the mass contrast between adjacent side facets forms domain walls at the hinges, which host gapless chiral modes propagating along the $z$ direction. However, the mass term vanishes at the $\Gamma$ point ($k_x=k_y=0$), leaving the $(001)$ surface Dirac cone intact. Thus $H(\bm k)$ exhibits gapless hinge states along the $z$-direction, while no horizontal chiral hinge states emerge for wires periodic along the $x-$ or $y$-direction.

To numerically verify the gapless hinge states, we consider a wire geometry with open boundary conditions along $x$ and $y$ and periodic boundary conditions along $z$. Under these boundary conditions, four in-gap states emerge for a given $k_z$, with each two degenerate, as shown in Fig.~\ref{HOTI_kz}(b). The twofold degeneracy reflects two diagonal hinge pairs related by crystalline symmetry, and the four in-gap states correspond to the four corners of the $x$-$y$ cross section. The probability densities of these four in gap state in the $x$-$y$ plane are concentrated on complementary diagonal corners [Figs.~\ref{HOTI_kz}(c) and \ref{HOTI_kz}(d)]. Together with the slopes of the in-gap dispersions, these real-space distributions determine the flow pattern of hinge states sketched in Fig.~\ref{HOTI_kz}(a). In Appendix \ref{appendixa}, we present the real-space tight-binding model used for numerical calculations.

\section{Chiral Majorana hinge modes}
\label{sec:hinge-majorana}

We further consider $s$-wave superconducting pairing based the model Hamiltonian $H$. In the Nambu basis, the Bogoliubov-de-Gennes (BdG) Hamiltonian is
\begin{align}
\mathcal{H}_{\mathrm{BdG}}(\mathbf{k})
=
\begin{pmatrix}
H(\mathbf{k})-\mu
&
\hat{\Delta}(\mathbf{k})
\\
\hat{\Delta}^{\dagger}(\mathbf{k})
&
-H^{*}(-\mathbf{k})+\mu
\end{pmatrix},
\label{eq:BdG-k-space}
\end{align}
where $\hat{\Delta}(\mathbf{k})=i\Delta\tau_0 \sigma_y$ denotes the onsite intraorbital $s$-wave pairing with $\Delta$ being the pairing amplitude, and $\mu$ is the chemical potential. The BdG Hamiltonian $\mathcal{H}_{\mathrm{BdG}}$ possesses an intrinsic particle-hole symmetry $\mathcal{P}=\rho_x\tau_0\sigma_0\mathcal{K}$, where $\rho_x$ acts on the particle-hole subspace, and therefore belongs to the Altland-Zirnbauer symmetry class D \cite{A-Z-class}. 
In Appendix \ref{appendixa}, we present the real-space tight-binding model of $\mathcal{H}_{\mathrm{BdG}}$ used for the numerical calculation.

We note that the $s$-wave superconducting pairing $\Delta$ gaps the gapless surface Dirac cone on the $(001)$ surface, whereas the side surfaces remain gapped by the magnetic mass $m$, which dominates over $\Delta$ in our setting. The coexistence of magnetic and superconducting gaps on different facets yields new mass domain walls at the interfaces between side facets (gapped by $m$) and the top/bottom facets (gapped by $\Delta$), which lead additional chiral Majorana hinge modes. Importantly, for the pre-existing gapless hinge states of the normal-state Hamiltonian, the superconducting pairing cannot open a gap because each hinge hosts only one propagating chiral mode. Consequently, these $z$-directed chiral hinge states persist as gapless Majorana hinge modes in the superconducting system.

\begin{figure}
	\includegraphics[width=8.5cm]{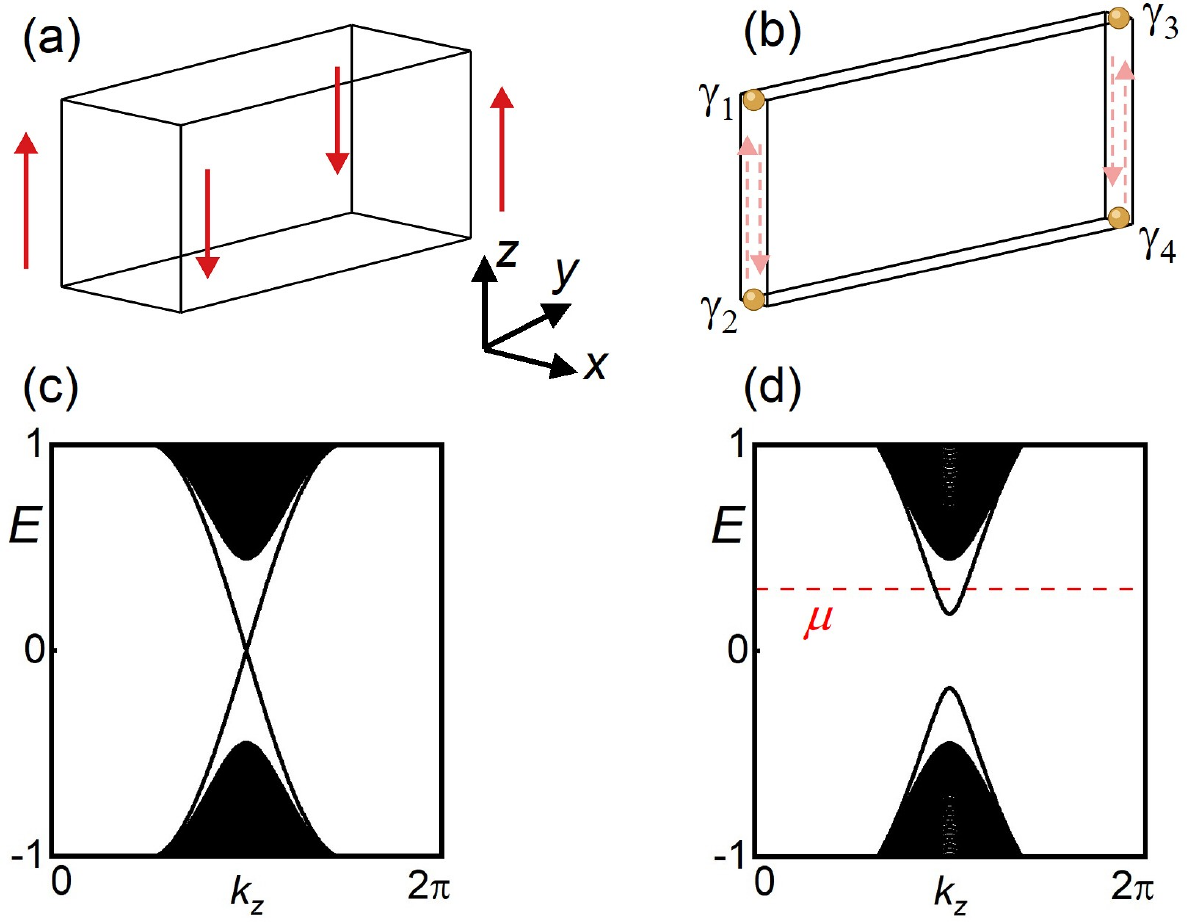}
\caption{Finite-size-induced reconstruction of the chiral HOTI hinge spectrum.  (a) Schematic illustration of decoupled chiral hinge states in the thick limit. (b) Schematic illustration of coupled hinge states forming helical channels in the thin limit; upon superconducting proximitization, these channels terminate in MZMs $\gamma_{1,2,3,4}$ at their endpoints. (c) Normal-state spectrum for $L_x=10$ and $L_y=80$, corresponding to the decoupled limit in (a). (d) Normal-state spectrum for $L_x=2$ and $L_y=80$, corresponding to the coupled limit in (b). The red dashed line marks a chemical potential within the  energy window of the coupled hinge states. The common model parameters are $t=1$, $M=2$, $\lambda=1$, and $m=0.5$.}
	\label{band_HOTI_mn}
\end{figure}

We numerically verify the existence of the chiral Majorana hinge modes in $\mathcal{H}_{\mathrm{BdG}}$. For wire geometries periodic along $z$, $x$, and $y$, the corresponding energy spectra are shown in Figs.~\ref{hinge_HOTSC}(b)--\ref{hinge_HOTSC}(d), respectively. In-gap states emerge under all these three boundary conditions. Specifically,  for the $z$-periodic wire, eight in-gap states appear at a given $k_z$, corresponding to two chiral Majorana hinge modes on each hinge that descend from one chiral hinge mode of the normal state. For the $x$-periodic ($y$-periodic) wire, four in-gap states appear at a given $k_x$ ($k_y$), associated with the four corners of the $x$-$z$ ($y$-$z$) cross section. Figs.~\ref{hinge_HOTSC}(e)--\ref{hinge_HOTSC}(g) plot the probability distributions of these in-gap Majorana modes marked in Figs.~\ref{hinge_HOTSC}(b)--\ref{hinge_HOTSC}(d), respectively.
Appendix~\ref{app:superconducting-spectra} further provides the real space distribution plot of the in-gap modes and we can identify the flow pattern of these chiral Majorana hinge modes, as sketched in Fig.~\ref{hinge_HOTSC}(a).

\section{Quasi-two-dimensional Majorana zero modes}
\label{sec:q2d-mzm}

\begin{figure*}
	\includegraphics[width=16cm]{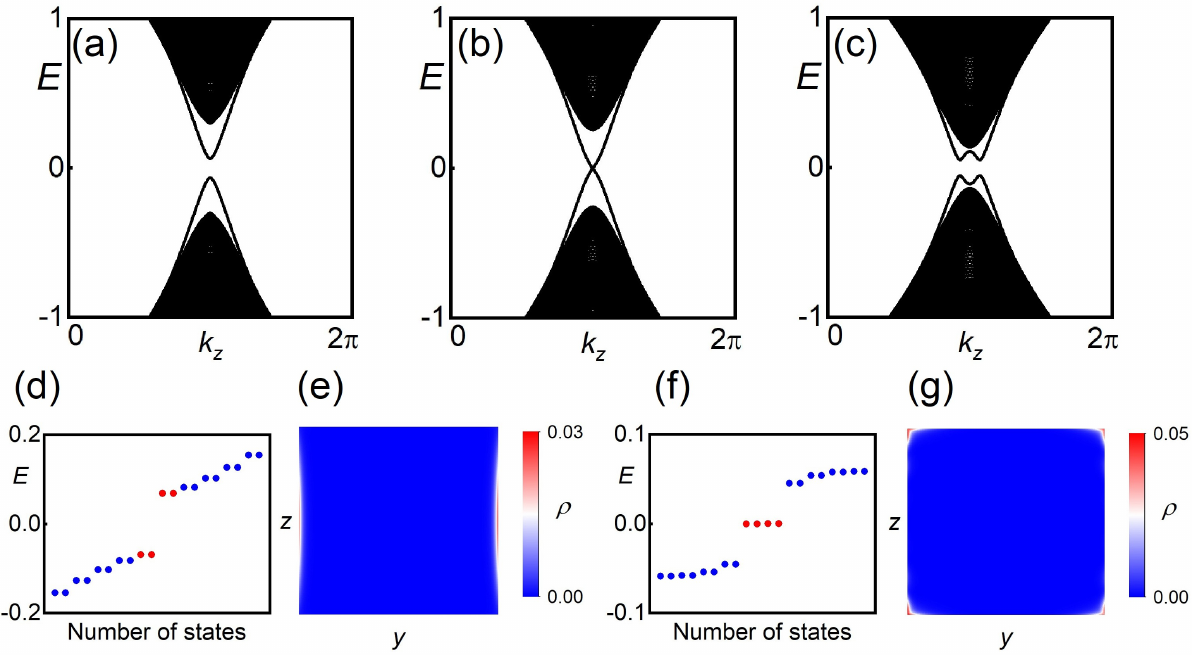}
\caption{Energy spectral evolution and real-space distribution of MZMs.  (a)--(c) BdG energy spectra with periodic boundary conditions along $z$ and open boundary conditions in the $x$-$y$ plane for a $2\times80$ cross section, at $\mu=0.1$, $0.17$, and $0.3$, respectively. The gap closes near the transition point (b) and reopens for larger $\mu$ (c). (d) Energy spectrum under open boundary conditions for the parameters in (a) with system size $2\times80\times80$. Red points mark the low-energy states used in (e). (e) Density projected along the thin $x$ direction for the marked states in (d). (f) Energy spectrum of under the open boundary conditions for the parameters in (c), where four MZMs appear inside the superconducting gap. (g) $x$-projected density of the MZMs in (f). The common parameters are $t=1$, $M=2$, $\lambda=1$, $m=0.5$, and $\Delta=0.1$.}

	\label{MZM}
\end{figure*}

Similar to chiral Majorana edge states, chiral Majorana hinge states are inherently delocalized and cannot be used for topological quantum computation. In the following, we utilize finite-size coupling of chiral hinge states to realize quasi-two-dimensional MZMs.

We first focus on the normal-state Hamiltonian $H$. We reduce the sample thickness along $x$ while keeping the other two directions sufficiently large, namely $L_y, L_z \gg L_x$. For a thick sample ($L_x \gg \xi$ with $\xi$ the hinge-state penetration depth), the $z$-directed chiral hinge states on opposite $x$ sides are effectively decoupled [Fig.~\ref{band_HOTI_mn}(a)], yielding gapless chiral modes as shown in Fig.~\ref{band_HOTI_mn}(c). When $L_x$ is reduced to a few lattice constants, the wave functions of opposite $z$-directed hinge states overlap substantially across the thin direction and hybridize [Fig.~\ref{band_HOTI_mn}(b)]. This finite-size hybridization opens a gap near $k_z=0$ and reconstructs the low-energy spectrum into two separate bands [Fig.~\ref{band_HOTI_mn}(d)]. Importantly, because the sample remains wide along $y$, the hinge states near the two opposite $y$ boundaries remain spatially well separated. Consequently, the gapless chiral hinge states along $z$ reorganize into two spatially separated one-dimensional channels, each localized near one of the two $y$ edges. Within each channel, the hybridization pairs the original right-moving hinge state from one $x$ side with the left-moving hinge state from the opposite $x$ side, forming an effective helical channel with counter-propagating modes. When the chemical potential is tuned into the finite-size-reconstructed energy window of hinge states, proximity-induced $s$-wave pairing projected onto these  helical channels opens an effective $p$-wave superconducting gap. These two hinge-derived helical channels then behave as effective $p$-wave superconducting wires, terminating in four localized MZMs when the $z$ direction is opened. The mechanism is closely analogous to the narrow-QAH construction, but the channels here are formed from chiral HOTI hinge states and remain separated near the two opposite $y$ boundaries.

The chemical potential determines whether the proximitized hinge-derived channels realize a topological superconducting phase. Figures~\ref{MZM}(a)--\ref{MZM}(c) show the evolution of the low-energy BdG spectrum at fixed pairing amplitude $\Delta$ but varying chemical potential $\mu$. As $\mu$ is tuned, a topological phase transition occurs, signaled by a gap closing and reopening [Fig.~\ref{MZM}(b)]. This transition separates the trivial phase [Fig.~\ref{MZM}(a)] from the topological phase [Fig.~\ref{MZM}(c)]. In the trivial phase, the chemical potential lies outside the gapped hinge-derived bands of the normal state, and the system exhibits a fully gapped excitation spectrum without in-gap states. The fully open boundary spectrum shows no zero modes [Fig.~\ref{MZM}(d)], and the real-space density projected along the thin $x$ direction---i.e., summed over the spatial coordinate $x$---does not exhibit four isolated peaks at the corners of the $y$-$z$ plane [Fig.~\ref{MZM}(e)], confirming the nontopological nature of this regime. In the topological phase, by contrast, the chemical potential crosses the gapped hinge-derived bands, and the proximity-induced pairing opens a topological gap. The fully open boundary spectrum then hosts four MZMs within the superconducting gap [Fig.~\ref{MZM}(f)]. These zero modes are localized at the four corners of the $y$-$z$ plane, as demonstrated by the projected density summed over the spatial coordinate $x$ [Fig.~\ref{MZM}(g)].

\begin{figure}
	\includegraphics[width=7cm]{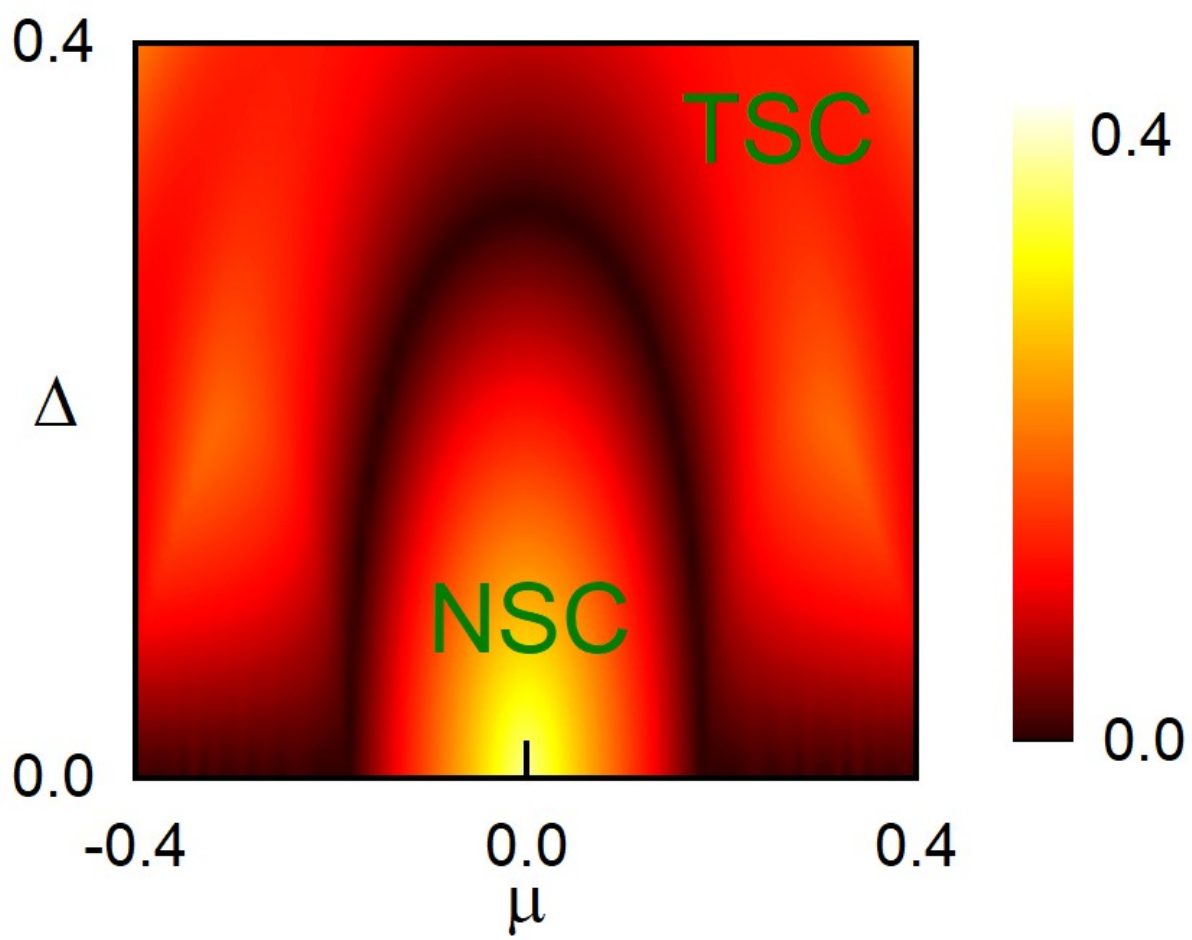}
	\caption{Phase diagram of the quasi-two-dimensional BdG system in the parameter space of the chemical potential $\mu$ and proximity-induced pairing amplitude $\Delta$. TSC denotes the topological superconducting regime in which the fully open geometry hosts four MZMs. NSC denotes the nontopological superconducting regime without MZMs. Dark gap-closing lines mark the phase boundaries.}
	\label{phase}
\end{figure}

Figure~\ref{phase} presents the phase diagram obtained from the minimum positive BdG energy gap as a function of $\mu$ and $\Delta$. The dark gap-closing lines mark the boundaries between the TSC regime, which hosts four MZMs, and the nontopological superconducting (NSC) regime without zero modes. The finite extent of the TSC region demonstrates that the MZMs require no fine tuning and persist over a broad parameter window.

\section{Robustness of MZMs to Disorder}
\label{sec:disorder}

\begin{figure*}
	\includegraphics[width=15cm]{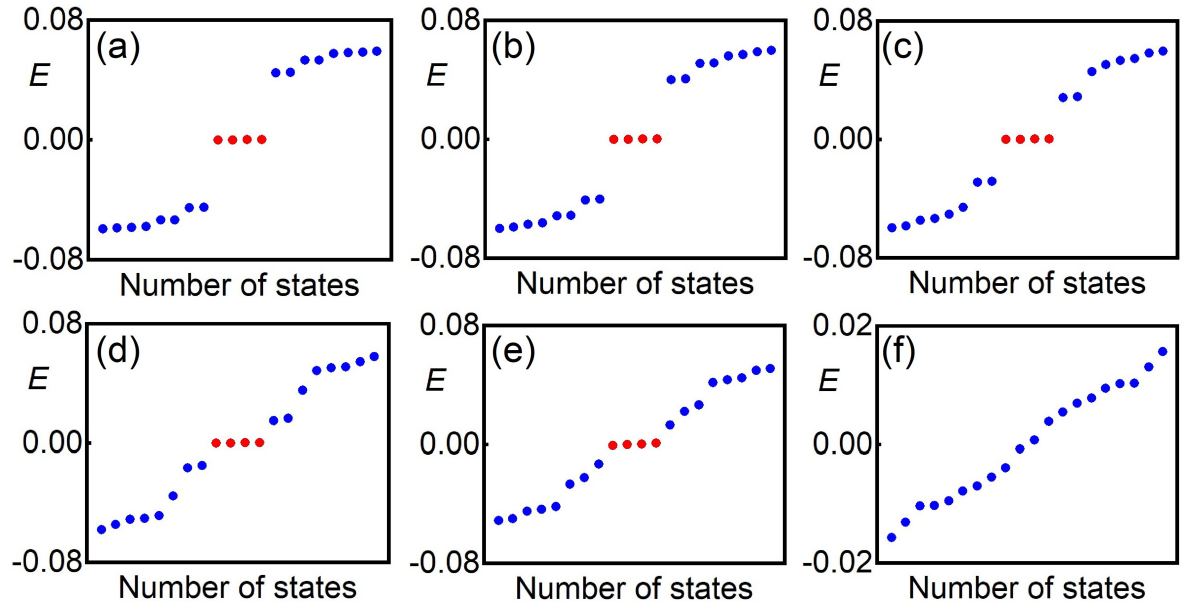}
	\caption{Representative low-energy spectra with random onsite potential $W_u$ and orbital-mass disorder $W_m$. Panels (a)--(f) correspond to $(W_u,W_m)=(0.05,0.05)$, $(0.2,0.1)$, $(0.4,0.2)$, $(0.6,0.3)$, $(1,0.5)$, and $(2,1)$, respectively. Red points in (a)--(e) identify the four MZMs that remain separated from finite-energy states by a nonzero excitation gap. At the strongest disorder in (f),  the zero-mode structure is no longer resolved.}
	\label{disorder}
\end{figure*}

We now study the disorder effect on the MZMs. We consider two types of nonmagnetic disorder: a random onsite potential and random orbital-mass fluctuations,
\begin{align}
 H_{\mathrm{dis}}=\sum_{\mathbf{r},\alpha=a,b}\sum_s\left[\delta u_{\mathbf{r}}+\eta_{\alpha}\delta m_{\mathbf{r}}\right]c_{\mathbf{r},\alpha,s}^{\dagger}c_{\mathbf{r},\alpha,s},
 \label{eq:disorder-real-space}
\end{align}
where $c_{\mathbf{r},\alpha,s}^{\dagger}$ creates an electron at site $\mathbf{r}$ in orbital $\alpha$ with spin $s$, and $\eta_a=+1$, $\eta_b=-1$. The random variables $\delta u_{\mathbf{r}}$ and $\delta m_{\mathbf{r}}$ are drawn from uniform distributions $[-W_u/2,W_u/2]$ and $[-W_m/2,W_m/2]$, respectively. The first term shifts the local chemical potential, while the second term modifies the band-inversion strength. In Nambu notation, these operators read $\rho_z\tau_0\sigma_0$ and $\rho_z\tau_z\sigma_0$, both preserving the particle-hole symmetry of the BdG Hamiltonian. A generic disorder realization breaks $\mathcal{C}_{4z}\mathcal{T}$ because symmetry-related sites carry uncorrelated random values.

The spectra in Fig.~\ref{disorder} demonstrate that the four MZMs are robust against weak-to-moderate disorder. In Figs.~\ref{disorder}(a)--\ref{disorder}(e), the zero modes remain within the excitation gap, separated from finite-energy states. Increasing $W_u$ strengthens local fluctuations of the Fermi level, while increasing $W_m$ perturbs the band-inversion mass that controls both the parent HOTI and the finite-size hinge hybridization. As long as the excitation gap remains finite, spatially separated MZMs at different boundary regions are not efficiently coupled by local disorder. At strong disorder, the gap is filled and the zero-mode structure is no longer resolved, which implies that the system is trivialized [Fig.~\ref{disorder}(f)].

\section{Conclusion and discussion}
\label{sec:conclusion}
In summary, we propose using finite-size-coupled chiral hinge states to realize spatially separated quasi-two-dimensional MZMs in a HOTI proximitized by a conventional $s$-wave superconductor. We demonstrate this proposal by studying a concrete theoretical model, establishing that the topological superconducting phase can be realized when the chemical potential is tuned into the finite-size-reconstructed hinge bands. Since chiral hinge states in magnetic topological insulators and conventional superconducting proximity effects are both experimentally accessible \cite{PhysRevLett.124.136407,PhysRevLett.122.256402,Yue2019,ZHAO20253310}, our scheme offers a feasible platform for realizing Majorana corner states. The resulting zero-bias conductance peaks could be detected by scanning tunneling microscopy, providing a clear spectroscopic signature of the topological Majorana corner modes \cite{LawKT2009}.

\section{ACKNOWLEDGMENTS}
We would like to thank Xilin Feng and Yingming Xie for useful discussions.

\begin{figure*}
	\includegraphics[width=16cm]{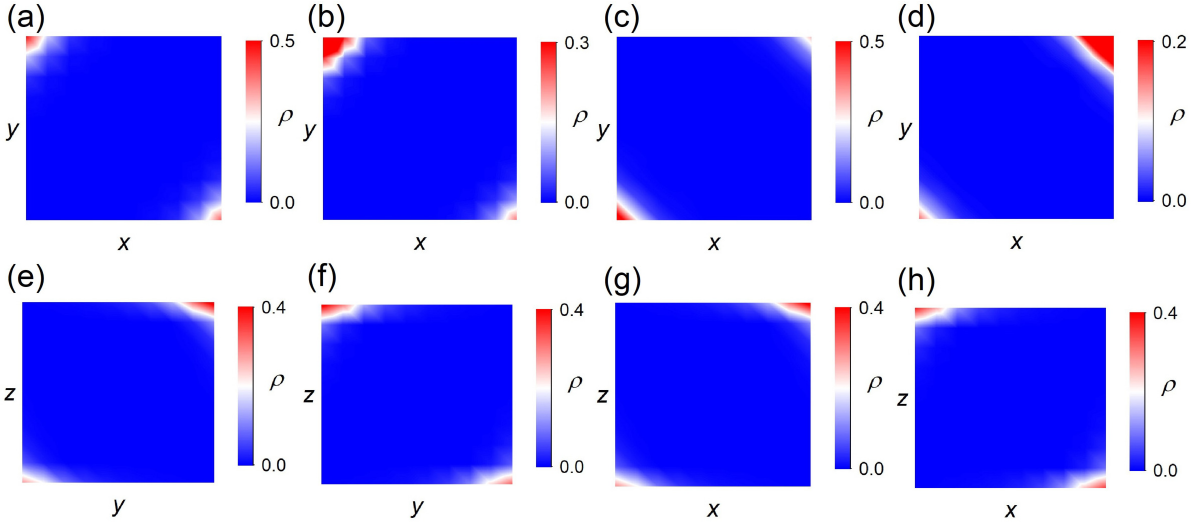}
	\caption{Individual probability densities of the in-gap Majorana hinge modes marked \textit{A}--\textit{H} in Figs.~\ref{hinge_HOTSC}(b)--\ref{hinge_HOTSC}(d). Panels (a)--(h) correspond to \textit{A}--\textit{H}, respectively.}
	\label{HOTSC_appendix-2}
\end{figure*}

\appendix

\section{ Real space tight-binding model}
\label{appendixa}
In this section, we present the real space tight-binding model on a cubic lattice.
At site $\mathbf r$, we use the basis $c_{\mathbf r}=(c_{\mathbf r,a,\uparrow},c_{\mathbf r,a,\downarrow},c_{\mathbf r,b,\uparrow},c_{\mathbf r,b,\downarrow})^T$, where $a$ and $b$ label two orbitals. Pauli matrices $\tau_i$ act in orbital space, while $\sigma_i$ act in spin space. The normal-state Hamiltonian is
\begin{align}
 H_{\mathrm C}&=H_M+H_t+H_{\lambda}+H_{m},
 \label{eq:normal-real-space}\\
 H_M&=M\sum_{\mathbf r}c_{\mathbf r}^{\dagger}\tau_z\sigma_0c_{\mathbf r},\notag\\
 H_t&=\frac{t}{2}\sum_{\mathbf r,i}\left(c_{\mathbf r+\hat e_i}^{\dagger}\tau_z\sigma_0c_{\mathbf r}+\mathrm{H.c.}\right),\notag\\
 H_{\lambda}&=\frac{\lambda}{2\mathrm{i}}\sum_{\mathbf r,i}\left(c_{\mathbf r+\hat e_i}^{\dagger}\tau_x\sigma_i c_{\mathbf r}-\mathrm{H.c.}\right),\notag\\
 H_{m}&=\frac{m}{2}\sum_{\mathbf r,i}n_i\left(c_{\mathbf r+\hat e_i}^{\dagger}\tau_y\sigma_0c_{\mathbf r}+\mathrm{H.c.}\right),\notag
\end{align}
where $i=x,y,z$, $(n_x,n_y,n_z)=(1,-1,0)$, and the lattice constant is set to unity. 

After considering conventional $s$-wave superconductor pairing, the BdG Hamiltonian is
\begin{align}
 H_{\mathrm{BdG}}=&H_{\mathrm C}
 -\mu\sum_{\mathbf r,\alpha=a,b}\sum_s c_{\mathbf r,\alpha,s}^{\dagger}c_{\mathbf r,\alpha,s}\nonumber\\
 &+\sum_{\mathbf r,\alpha=a,b}\left(\Delta c_{\mathbf r,\alpha,\uparrow}^{\dagger}c_{\mathbf r,\alpha,\downarrow}^{\dagger}+\mathrm{H.c.}\right).
 \label{eq:SC-real-space}
\end{align}
Here $\mu$ is the chemical potential, and $\Delta$ is the superconductor pairing amplitude.

\section{Real space distribution of Majorana hinge modes}
\label{app:superconducting-spectra}

In this section, we present the detailed real-space distributions of the chiral Majorana hinge modes shown in Fig.~\ref{hinge_HOTSC}. For wire geometries periodic along $z$, $x$, and $y$, we label representative in-gap states as $A$--$D$, $E$--$F$, and $G$--$H$ at fixed $k_z$, $k_x$, and $k_y$, respectively. Their real-space probability densities are shown in Figs.~\ref{HOTSC_appendix-2}(a)--\ref{HOTSC_appendix-2}(h), respectively. We can see that every individual in-gap state is distributed at the corresponding corner, which demonstrates the propagating channels of the Majorana hinge modes  sketched in Fig.~\ref{hinge_HOTSC} (a).

\bibliography{references}

@article{Kitaev2001,
  author = {A. Yu. Kitaev},
  title = {Unpaired Majorana fermions in quantum wires},
  journal = {Phys. Usp.},
  volume = {44},
  pages = {131--136},
  year = {2001},
  doi = {10.1070/1063-7869/44/10S/S29},
  url = {https://doi.org/10.1070/1063-7869/44/10S/S29}
}

@article{Nayak2008RMP,
  author = {C. Nayak and S. H. Simon and A. Stern and M. Freedman and S. Das Sarma},
  title = {Non-Abelian anyons and topological quantum computation},
  journal = {Rev. Mod. Phys.},
  volume = {80},
  pages = {1083--1159},
  year = {2008},
  doi = {10.1103/RevModPhys.80.1083},
  url = {https://doi.org/10.1103/RevModPhys.80.1083}
}

@article{Alicea2012RPP,
  author = {J. Alicea},
  title = {New directions in the pursuit of Majorana fermions in solid state systems},
  journal = {Rep. Prog. Phys.},
  volume = {75},
  pages = {076501},
  year = {2012},
  doi = {10.1088/0034-4885/75/7/076501},
  url = {https://doi.org/10.1088/0034-4885/75/7/076501}
}

@article{PhysRevB.108.075143,
  title = {Generalization of Benalcazar-Bernevig-Hughes model to arbitrary dimensions},
  author = {Luo, Xun-Jiang and Wu, Fengcheng},
  journal = {Phys. Rev. B},
  volume = {108},
  issue = {7},
  pages = {075143},
  numpages = {10},
  year = {2023},
  month = {Aug},
  publisher = {American Physical Society},
  doi = {10.1103/PhysRevB.108.075143},
  url = {https://link.aps.org/doi/10.1103/PhysRevB.108.075143}
}

@article{LawKT2009,
  title = {Majorana Fermion Induced Resonant Andreev Reflection},
  author = {Law, K. T. and Lee, Patrick A. and Ng, T. K.},
  journal = {Phys. Rev. Lett.},
  volume = {103},
  issue = {23},
  pages = {237001},
  numpages = {4},
  year = {2009},
  month = {Dec},
  publisher = {American Physical Society},
  doi = {10.1103/PhysRevLett.103.237001},
  url = {https://link.aps.org/doi/10.1103/PhysRevLett.103.237001}
}

@article{ZHAO20253310,
title = {Revealing higher-order topological bulk-boundary correspondence in bismuth crystal with spin-helical hinge state loop and proximity superconductivity},
journal = {Science Bulletin},
volume = {70},
number = {20},
pages = {3310-3313},
year = {2025},
issn = {2095-9273},
doi = {https://doi.org/10.1016/j.scib.2025.08.047},
url = {https://www.sciencedirect.com/science/article/pii/S2095927325008783},
author = {Dongming Zhao and Yang Zhong and Tian Yuan and Haitao Wang and Tianxing Jiang and Yang Qi and Hongjun Xiang and Xingao Gong and Donglai Feng and Tong Zhang}
}

@article{PhysRevB.107.045118,
  title = {Higher-order topological phases emerging from Su-Schrieffer-Heeger stacking},
  author = {Luo, Xun-Jiang and Pan, Xiao-Hong and Liu, Chao-Xing and Liu, Xin},
  journal = {Phys. Rev. B},
  volume = {107},
  issue = {4},
  pages = {045118},
  numpages = {14},
  year = {2023},
  month = {Jan},
  publisher = {American Physical Society},
  doi = {10.1103/PhysRevB.107.045118},
  url = {https://link.aps.org/doi/10.1103/PhysRevB.107.045118}
}

@Article{Yue2019,
author={Yue, Changming
and Xu, Yuanfeng
and Song, Zhida
and Weng, Hongming
and Lu, Yuan-Ming
and Fang, Chen
and Dai, Xi},
title={{Symmetry-enforced chiral hinge states and surface quantum anomalous Hall effect in the magnetic axion insulator $\text{Bi}_{2-x}\text{Sm}_x\text{Se}_3$}},
journal={Nat. Phys.},
year={2019},
month={Jun},
day={01},
volume={15},
number={6},
pages={577-581},
issn={1745-2481},
doi={10.1038/s41567-019-0457-0},
url={https://doi.org/10.1038/s41567-019-0457-0}
}

@article{PhysRevLett.124.136407,
  title = {{M\"obius Insulator and Higher-Order Topology in ${\mathrm{MnBi}}_{2n}{\mathrm{Te}}_{3n+1}$}},
  author = {Zhang, Rui-Xing and Wu, Fengcheng and Das Sarma, S.},
  journal = {Phys. Rev. Lett.},
  volume = {124},
  issue = {13},
  pages = {136407},
  numpages = {6},
  year = {2020},
  month = {Apr},
  publisher = {American Physical Society},
  doi = {10.1103/PhysRevLett.124.136407},
  url = {https://link.aps.org/doi/10.1103/PhysRevLett.124.136407}
}

@article{PhysRevLett.122.256402,
  title = {{Higher-Order Topology of the Axion Insulator ${\mathrm{EuIn}}_{2}{\mathrm{As}}_{2}$}},
  author = {Xu, Yuanfeng and Song, Zhida and Wang, Zhijun and Weng, Hongming and Dai, Xi},
  journal = {Phys. Rev. Lett.},
  volume = {122},
  issue = {25},
  pages = {256402},
  numpages = {6},
  year = {2019},
  month = {Jun},
  publisher = {American Physical Society},
  doi = {10.1103/PhysRevLett.122.256402},
  url = {https://link.aps.org/doi/10.1103/PhysRevLett.122.256402}
}

@article{Beenakker2013ARCMP,
  author = {C. W. J. Beenakker},
  title = {Search for Majorana fermions in superconductors},
  journal = {Annu. Rev. Condens. Matter Phys.},
  volume = {4},
  pages = {113--136},
  year = {2013},
  doi = {10.1146/annurev-conmatphys-030212-184337},
  url = {https://doi.org/10.1146/annurev-conmatphys-030212-184337}
}

@article{DasSarma2015NPJQI,
  author = {S. Das Sarma and M. Freedman and C. Nayak},
  title = {Majorana zero modes and topological quantum computation},
  journal = {npj Quantum Inf.},
  volume = {1},
  pages = {15001},
  year = {2015},
  doi = {10.1038/npjqi.2015.1},
  url = {https://doi.org/10.1038/npjqi.2015.1}
}

@article{SatoAndo2017RPP,
  author = {M. Sato and Y. Ando},
  title = {Topological superconductors: a review},
  journal = {Rep. Prog. Phys.},
  volume = {80},
  pages = {076501},
  year = {2017},
  doi = {10.1088/1361-6633/aa6ac7},
  url = {https://doi.org/10.1088/1361-6633/aa6ac7}
}

@article{Flensberg2021NatRevMater,
  author = {K. Flensberg and F. von Oppen and A. Stern},
  title = {Engineered platforms for topological superconductivity and Majorana zero modes},
  journal = {Nat. Rev. Mater.},
  volume = {6},
  pages = {944--958},
  year = {2021},
  doi = {10.1038/s41578-021-00336-6},
  url = {https://doi.org/10.1038/s41578-021-00336-6}
}

@article{MackenzieMaeno2003,
  author = {A. P. Mackenzie and Y. Maeno},
  title = {{The superconductivity of Sr$_2$RuO$_4$ and the physics of spin-triplet pairing}},
  journal = {Rev. Mod. Phys.},
  volume = {75},
  pages = {657--712},
  year = {2003},
  doi = {10.1103/RevModPhys.75.657},
  url = {https://doi.org/10.1103/RevModPhys.75.657}
}

@article{FuKane2008PRL,
  author = {L. Fu and C. L. Kane},
  title = {Superconducting proximity effect and Majorana fermions at the surface of a topological insulator},
  journal = {Phys. Rev. Lett.},
  volume = {100},
  pages = {096407},
  year = {2008},
  doi = {10.1103/PhysRevLett.100.096407},
  url = {https://doi.org/10.1103/PhysRevLett.100.096407}
}

@article{Sau2010PRL,
  author = {J. D. Sau and R. M. Lutchyn and S. Tewari and S. Das Sarma},
  title = {Generic new platform for topological quantum computation using semiconductor heterostructures},
  journal = {Phys. Rev. Lett.},
  volume = {104},
  pages = {040502},
  year = {2010},
  doi = {10.1103/PhysRevLett.104.040502},
  url = {https://doi.org/10.1103/PhysRevLett.104.040502}
}

@article{Lutchyn2010PRL,
  author = {R. M. Lutchyn and J. D. Sau and S. Das Sarma},
  title = {Majorana fermions and a topological phase transition in semiconductor-superconductor heterostructures},
  journal = {Phys. Rev. Lett.},
  volume = {105},
  pages = {077001},
  year = {2010},
  doi = {10.1103/PhysRevLett.105.077001},
  url = {https://doi.org/10.1103/PhysRevLett.105.077001}
}

@article{Oreg2010PRL,
  author = {Y. Oreg and G. Refael and F. von Oppen},
  title = {Helical liquids and Majorana bound states in quantum wires},
  journal = {Phys. Rev. Lett.},
  volume = {105},
  pages = {177002},
  year = {2010},
  doi = {10.1103/PhysRevLett.105.177002},
  url = {https://doi.org/10.1103/PhysRevLett.105.177002}
}

@article{Mourik2012Science,
  author = {V. Mourik and K. Zuo and S. M. Frolov and S. R. Plissard and E. P. A. M. Bakkers and L. P. Kouwenhoven},
  title = {Signatures of Majorana fermions in hybrid superconductor-semiconductor nanowire devices},
  journal = {Science},
  volume = {336},
  pages = {1003--1007},
  year = {2012},
  doi = {10.1126/science.1222360},
  url = {https://doi.org/10.1126/science.1222360}
}

@article{NadjPerge2014Science,
  author = {S. Nadj-Perge and I. K. Drozdov and J. Li and H. Chen and S. Jeon and J. Seo and A. H. MacDonald and B. A. Bernevig and A. Yazdani},
  title = {Observation of Majorana fermions in ferromagnetic atomic chains on a superconductor},
  journal = {Science},
  volume = {346},
  pages = {602--607},
  year = {2014},
  doi = {10.1126/science.1259327},
  url = {https://doi.org/10.1126/science.1259327}
}

@article{Lutchyn2018NatRevMater,
  author = {R. M. Lutchyn and E. P. A. M. Bakkers and L. P. Kouwenhoven and P. Krogstrup and C. M. Marcus and Y. Oreg},
  title = {Majorana zero modes in superconductor-semiconductor heterostructures},
  journal = {Nat. Rev. Mater.},
  volume = {3},
  pages = {52--68},
  year = {2018},
  doi = {10.1038/s41578-018-0003-1},
  url = {https://doi.org/10.1038/s41578-018-0003-1}
}

@article{Deng2018Nonlocality,
  author = {M.-T. Deng and S. Vaitiek{\=e}nas and E. Prada and P. San-Jose and J. Nyg{\aa}rd and P. Krogstrup and R. Aguado and C. M. Marcus},
  title = {Nonlocality of Majorana modes in hybrid nanowires},
  journal = {Phys. Rev. B},
  volume = {98},
  pages = {085125},
  year = {2018},
  doi = {10.1103/PhysRevB.98.085125},
  url = {https://doi.org/10.1103/PhysRevB.98.085125}
}

@article{Wang2018FeTeSe,
  author = {D. Wang and L. Kong and P. Fan and H. Chen and S. Zhu and W. Liu and L. Cao and Y. Sun and S. Du and J. Schneeloch and R. Zhong and G. Gu and L. Fu and H. Ding and H.-J. Gao},
  title = {Evidence for Majorana bound states in an iron-based superconductor},
  journal = {Science},
  volume = {362},
  pages = {333--335},
  year = {2018},
  doi = {10.1126/science.aao1797},
  url = {https://doi.org/10.1126/science.aao1797}
}

@article{Machida2019NatMater,
  author = {T. Machida and Y. Sun and S. Pyon and S. Takeda and Y. Kohsaka and T. Hanaguri and T. Sasagawa and T. Tamegai},
  title = {Zero-energy vortex bound state in the superconducting topological surface state of Fe(Se,Te)},
  journal = {Nat. Mater.},
  volume = {18},
  pages = {811--815},
  year = {2019},
  doi = {10.1038/s41563-019-0397-1},
  url = {https://doi.org/10.1038/s41563-019-0397-1}
}

@article{Wang2015QAHSC,
  author = {J. Wang and Q. Zhou and B. Lian and S.-C. Zhang},
  title = {Chiral topological superconductor and half-integer conductance plateau from quantum anomalous Hall plateau transition},
  journal = {Phys. Rev. B},
  volume = {92},
  pages = {064520},
  year = {2015},
  doi = {10.1103/PhysRevB.92.064520},
  url = {https://doi.org/10.1103/PhysRevB.92.064520}
}

@article{Chen2018QAHMajorana,
  author = {C.-Z. Chen and Y.-M. Xie and J. Liu and P. A. Lee and K. T. Law},
  title = {Quasi-one-dimensional quantum anomalous Hall systems as new platforms for scalable topological quantum computation},
  journal = {Phys. Rev. B},
  volume = {97},
  pages = {104504},
  year = {2018},
  doi = {10.1103/PhysRevB.97.104504},
  url = {https://doi.org/10.1103/PhysRevB.97.104504}
}

@article{Zeng2018,
  author = {Yongxin Zeng and Chao Lei and Gaurav Chaudhary and Allan H. MacDonald},
  title = {Quantum anomalous Hall Majorana platform},
  journal = {Phys. Rev. B},
  volume = {97},
  pages = {081102},
  year = {2018},
  doi = {10.1103/PhysRevB.97.081102},
  url = {https://doi.org/10.1103/PhysRevB.97.081102}
}

@article{ShiozakiSato2014PRB,
  author = {K. Shiozaki and M. Sato},
  title = {Topology of crystalline insulators and superconductors},
  journal = {Phys. Rev. B},
  volume = {90},
  pages = {165114},
  year = {2014},
  doi = {10.1103/PhysRevB.90.165114},
  url = {https://doi.org/10.1103/PhysRevB.90.165114}
}

@article{Benalcazar2017Science,
  author = {W. A. Benalcazar and B. A. Bernevig and T. L. Hughes},
  title = {Quantized electric multipole insulators},
  journal = {Science},
  volume = {357},
  pages = {61--66},
  year = {2017},
  doi = {10.1126/science.aah6442},
  url = {https://doi.org/10.1126/science.aah6442}
}

@article{Benalcazar2017PRB,
  author = {W. A. Benalcazar and B. A. Bernevig and T. L. Hughes},
  title = {Electric multipole moments, topological multipole moment pumping, and chiral hinge states in crystalline insulators},
  journal = {Phys. Rev. B},
  volume = {96},
  pages = {245115},
  year = {2017},
  doi = {10.1103/PhysRevB.96.245115},
  url = {https://doi.org/10.1103/PhysRevB.96.245115}
}

@article{Langbehn2017PRL,
  author = {J. Langbehn and Y. Peng and L. Trifunovic and F. von Oppen and P. W. Brouwer},
  title = {Reflection-symmetric second-order topological insulators and superconductors},
  journal = {Phys. Rev. Lett.},
  volume = {119},
  pages = {246401},
  year = {2017},
  doi = {10.1103/PhysRevLett.119.246401},
  url = {https://doi.org/10.1103/PhysRevLett.119.246401}
}

@article{Song2017PRL,
  author = {Z. Song and Z. Fang and C. Fang},
  title = {$(d-2)$-dimensional edge states of rotation-symmetry-protected topological states},
  journal = {Phys. Rev. Lett.},
  volume = {119},
  pages = {246402},
  year = {2017},
  doi = {10.1103/PhysRevLett.119.246402},
  url = {https://doi.org/10.1103/PhysRevLett.119.246402}
}

@article{Schindler2018SciAdv,
  author = {F. Schindler and A. M. Cook and M. G. Vergniory and Z. Wang and S. S. P. Parkin and B. A. Bernevig and T. Neupert},
  title = {Higher-order topological insulators},
  journal = {Sci. Adv.},
  volume = {4},
  pages = {eaat0346},
  year = {2018},
  doi = {10.1126/sciadv.aat0346},
  url = {https://doi.org/10.1126/sciadv.aat0346}
}

@article{Khalaf2018PRB,
  author = {E. Khalaf},
  title = {Higher-order topological insulators and superconductors protected by inversion symmetry},
  journal = {Phys. Rev. B},
  volume = {97},
  pages = {205136},
  year = {2018},
  doi = {10.1103/PhysRevB.97.205136},
  url = {https://doi.org/10.1103/PhysRevB.97.205136}
}

@article{Geier2018PRB,
  author = {M. Geier and L. Trifunovic and M. Hoskam and P. W. Brouwer},
  title = {Second-order topological insulators and superconductors with an order-two crystalline symmetry},
  journal = {Phys. Rev. B},
  volume = {97},
  pages = {205135},
  year = {2018},
  doi = {10.1103/PhysRevB.97.205135},
  url = {https://doi.org/10.1103/PhysRevB.97.205135}
}

@article{Fu2021PRResearch,
  author = {B. Fu and Z.-A. Hu and S.-Q. Shen},
  title = {Bulk-hinge correspondence and three-dimensional quantum anomalous Hall effect in second-order topological insulators},
  journal = {Phys. Rev. Research},
  volume = {3},
  pages = {033177},
  year = {2021},
  doi = {10.1103/PhysRevResearch.3.033177},
  url = {https://doi.org/10.1103/PhysRevResearch.3.033177}
}

@article{WangLinHughes2018WeakPairing,
  author = {Y. Wang and M. Lin and T. L. Hughes},
  title = {Weak-pairing higher order topological superconductors},
  journal = {Phys. Rev. B},
  volume = {98},
  pages = {165144},
  year = {2018},
  doi = {10.1103/PhysRevB.98.165144},
  url = {https://doi.org/10.1103/PhysRevB.98.165144}
}

@article{Hsu2018HOTIMajorana,
  author = {C.-H. Hsu and P. Stano and J. Klinovaja and D. Loss},
  title = {Majorana Kramers pairs in higher-order topological insulators},
  journal = {Phys. Rev. Lett.},
  volume = {121},
  pages = {196801},
  year = {2018},
  doi = {10.1103/PhysRevLett.121.196801},
  url = {https://doi.org/10.1103/PhysRevLett.121.196801}
}

@article{Yan2018HighTPlatform,
  author = {Z. Yan and F. Song and Z. Wang},
  title = {Majorana corner modes in a high-temperature platform},
  journal = {Phys. Rev. Lett.},
  volume = {121},
  pages = {096803},
  year = {2018},
  doi = {10.1103/PhysRevLett.121.096803},
  url = {https://doi.org/10.1103/PhysRevLett.121.096803}
}

@article{Wang2018HighTempCorner,
  author = {Q. Wang and C.-C. Liu and Y.-M. Lu and F. Zhang},
  title = {High-temperature Majorana corner states},
  journal = {Phys. Rev. Lett.},
  volume = {121},
  pages = {186801},
  year = {2018},
  doi = {10.1103/PhysRevLett.121.186801},
  url = {https://doi.org/10.1103/PhysRevLett.121.186801}
}

@article{Yan2019SOTISCHeterostructure,
  author = {Z. Yan},
  title = {Majorana corner and hinge modes in second-order topological insulator/superconductor heterostructures},
  journal = {Phys. Rev. B},
  volume = {100},
  pages = {205406},
  year = {2019},
  doi = {10.1103/PhysRevB.100.205406},
  url = {https://doi.org/10.1103/PhysRevB.100.205406}
}

@article{Ghorashi2019SecondOrderDirac,
  author = {S. A. A. Ghorashi and X. Hu and T. L. Hughes and E. Rossi},
  title = {Second-order Dirac superconductors and magnetic field induced Majorana hinge modes},
  journal = {Phys. Rev. B},
  volume = {100},
  pages = {020509(R)},
  year = {2019},
  doi = {10.1103/PhysRevB.100.020509},
  url = {https://doi.org/10.1103/PhysRevB.100.020509}
}

@article{Ghorashi2020VortexSurfaceHOTI,
  author = {S. A. A. Ghorashi and T. L. Hughes and E. Rossi},
  title = {Vortex and surface phase transitions in superconducting higher-order topological insulators},
  journal = {Phys. Rev. Lett.},
  volume = {125},
  pages = {037001},
  year = {2020},
  doi = {10.1103/PhysRevLett.125.037001},
  url = {https://doi.org/10.1103/PhysRevLett.125.037001}
}

@article{Pan2019LatticeSOTSC,
  author = {X.-H. Pan and K.-J. Yang and L. Chen and G. Xu and C.-X. Liu and X. Liu},
  title = {Lattice-symmetry-assisted second-order topological superconductors and Majorana patterns},
  journal = {Phys. Rev. Lett.},
  volume = {123},
  pages = {156801},
  year = {2019},
  doi = {10.1103/PhysRevLett.123.156801},
  url = {https://doi.org/10.1103/PhysRevLett.123.156801}
}

@article{Zhu2019SecondOrderTSC,
  author = {X. Zhu},
  title = {Second-order topological superconductors with mixed pairing},
  journal = {Phys. Rev. Lett.},
  volume = {122},
  pages = {236401},
  year = {2019},
  doi = {10.1103/PhysRevLett.122.236401},
  url = {https://doi.org/10.1103/PhysRevLett.122.236401}
}

@article{Wu2020ZeemanSOTSC,
  author = {Y.-J. Wu and J. Hou and Y.-M. Li and X.-W. Luo and X. Shi and C. Zhang},
  title = {In-plane Zeeman-field-induced Majorana corner and hinge modes in an $s$-wave superconductor heterostructure},
  journal = {Phys. Rev. Lett.},
  volume = {124},
  pages = {227001},
  year = {2020},
  doi = {10.1103/PhysRevLett.124.227001},
  url = {https://doi.org/10.1103/PhysRevLett.124.227001}
}

@article{Kheirkhah2020CornerFlatBands,
  author = {M. Kheirkhah and Y. Nagai and C. Chen and F. Marsiglio},
  title = {Majorana corner flat bands in two-dimensional second-order topological superconductors},
  journal = {Phys. Rev. B},
  volume = {101},
  pages = {104502},
  year = {2020},
  doi = {10.1103/PhysRevB.101.104502},
  url = {https://doi.org/10.1103/PhysRevB.101.104502}
}

@article{ZhangColeDasSarma2019PRL,
  author = {R.-X. Zhang and W. S. Cole and S. Das Sarma},
  title = {Helical hinge Majorana modes in iron-based superconductors},
  journal = {Phys. Rev. Lett.},
  volume = {122},
  pages = {187001},
  year = {2019},
  doi = {10.1103/PhysRevLett.122.187001},
  url = {https://doi.org/10.1103/PhysRevLett.122.187001}
}

@article{Ghosh2021HierarchyHOTSC,
  author = {A. K. Ghosh and T. Nag and A. Saha},
  title = {Hierarchy of higher-order topological superconductors in three dimensions},
  journal = {Phys. Rev. B},
  volume = {104},
  pages = {134508},
  year = {2021},
  doi = {10.1103/PhysRevB.104.134508},
  url = {https://doi.org/10.1103/PhysRevB.104.134508}
}

@article{Chew2023TBGHOTSC,
  author = {A. Chew and Y. Wang and B. A. Bernevig and Z.-D. Song},
  title = {Higher-order topological superconductivity in twisted bilayer graphene},
  journal = {Phys. Rev. B},
  volume = {107},
  pages = {094512},
  year = {2023},
  doi = {10.1103/PhysRevB.107.094512},
  url = {https://doi.org/10.1103/PhysRevB.107.094512}
}

@article{Nakai2023HOTSCMultilayer,
  author = {R. Nakai and K. Nomura},
  title = {Higher-order topological superconductor phases in a multilayer system},
  journal = {Phys. Rev. B},
  volume = {108},
  pages = {184517},
  year = {2023},
  doi = {10.1103/PhysRevB.108.184517},
  url = {https://doi.org/10.1103/PhysRevB.108.184517}
}

@article{Luo2021WeakTIHOTSC,
  author = {X.-J. Luo and X.-H. Pan and X. Liu},
  title = {Higher-order topological superconductors based on weak topological insulators},
  journal = {Phys. Rev. B},
  volume = {104},
  pages = {104510},
  year = {2021},
  doi = {10.1103/PhysRevB.104.104510},
  url = {https://doi.org/10.1103/PhysRevB.104.104510}
}

@article{Pan2022BraidingMCM,
  author = {X.-H. Pan and X.-J. Luo and J.-H. Gao and X. Liu},
  title = {Detecting and braiding higher-order Majorana corner states through their spin degree of freedom},
  journal = {Phys. Rev. B},
  volume = {105},
  pages = {195106},
  year = {2022},
  doi = {10.1103/PhysRevB.105.195106},
  url = {https://doi.org/10.1103/PhysRevB.105.195106}
}

@article{Luo2024SpinBott,
  author = {Xun-Jiang Luo and Jia-Zheng Li and Meng Xiao and Fengcheng Wu},
  title = {Characterization of higher-order topological superconductors using {Bott} indices},
  journal = {Phys. Rev. B},
  volume = {111},
  pages = {184516},
  year = {2025},
  doi = {10.1103/PhysRevB.111.184516},
  url = {https://doi.org/10.1103/PhysRevB.111.184516}
}

@article{A-Z-class,
  title = {Classification of topological insulators and superconductors in three spatial dimensions},
  author = {Schnyder, Andreas P. and Ryu, Shinsei and Furusaki, Akira and Ludwig, Andreas W. W.},
  journal = {Phys. Rev. B},
  volume = {78},
  issue = {19},
  pages = {195125},
  numpages = {22},
  year = {2008},
  month = {Nov},
  publisher = {American Physical Society},
  doi = {10.1103/PhysRevB.78.195125},
  url = {https://link.aps.org/doi/10.1103/PhysRevB.78.195125}
}

\end{document}